\documentclass[twocolumn,showpacs,aps,pra,amsmath,amssymb]{revtex4}
\usepackage{bm,color,bbm}
\usepackage{hyperref, mathtools,graphicx}

\newcommand{\beq}{\begin{equation}}
\newcommand{\eeq}{\end{equation}}
\newcommand{\bqa}{\begin{eqnarray}}
\newcommand{\eqa}{\end{eqnarray}}
\newcommand{\nn}{\nonumber}

\newcommand{\erf}[1]{Eq.~(\ref{#1})}

\newcommand{\id}{\mathbbm{1}}

\definecolor{maroon}{rgb}{0.7,0,0}

\definecolor{ngreen}{rgb}{0.3,0.7,0.3}

\definecolor{golden}{rgb}{0.8,0.6,0.1}

\begin{document}

\title{Entanglement verification and steering when Alice and Bob cannot be trusted} 
\author{Eric  G. Cavalcanti$^{1,2}$, Michael J. W. Hall$^3$ and Howard M. Wiseman$^3$}
\affiliation{${}^1$School of Physics, University of Sydney, NSW 2006, Australia\\
${}^2$Quantum Group, Department of Computer Science, University of Oxford, OX1 3QD, United Kingdom\\
${}^3$Centre for Quantum Computation and Communication Technology (Australian Research Council), Centre for Quantum Dynamics, Griffith University, Brisbane, QLD 4111, Australia}
\begin{abstract}
Various protocols exist by which a referee can be convinced that two observers share an entangled resource.  Such protocols typically specify the types of communication allowed, and the degrees of trust required, between the referee and each observer.  Here it is shown that the need for any degree of trust of the observers by the referee can be completely removed via the referee using classical and quantum communication channels appropriately. In particular, trust-free verification of Bell nonlocality, EPR-steering, and entanglement, respectively, requires two classical channels, one classical and one quantum channel, and two quantum channels.  These channels correspond to suitable inputs of quantum randomness by the referee, which prevent the observers from mimicking entanglement using shared classical randomness. Our results generalize recent work by Buscemi [Phys. Rev. Lett. {\bf 108}, 200401 (2012)], and offer a perspective on the operational significance of that work. They also offer the possibility of simpler experimental demonstrations of the basic idea of quantum-refereed nonlocality tests.
\end{abstract}

\maketitle


\section{Introduction} \label{Sec:Intro}

Quantum entanglement is a remarkable and nonintuitive phenomenon, with no parallel in classical physics.  It also provides much more than a philosophical conversation piece --- various types of entanglement have been shown to provide useful physical resources \cite{frontier}, for tasks ranging from secure key generation \cite{secure} to distinguishing between two quantum channels \cite{piani}.

For two observers to convince a referee that they share an entangled resource, they must demonstrate a real physical effect that could not be achieved otherwise.  If the referee does not have direct access to the resource, and does not trust the observers, this is a nontrivial task --- a protocol is required that rules out all other possible explanations of the effect, and in particular rules out generation of the effect by purely local classical means. 

For example, it is clearly insufficient for the observers, after being isolated from each other, to each transmit a list of local measurement settings and measurement outcomes to the referee.  Even if two such lists, when combined, violate a Bell inequality \cite{bell}, they could simply have been generated before isolation via a conspiracy by the observers \cite{frontier}.  

To prevent the possibility of such conspiracies misleading the referee, it is necessary to have a protocol that negates the effects of any preexisting classical shared randomness between the observers. As will be shown here, this can always be done via a suitable injection of quantum randomness by the referee.  This is a consequence of, and offers an operational perspective on, recent work by Buscemi \cite{buscemi}.

In particular, Buscemi has shown that entanglement can always be witnessed by a suitable `semiquantum' game, involving one-way quantum channels from the referee to the observers. Nonorthogonality of the quantum states sent by the referee, via these channels, provides the necessary quantum randomness to ensure that the observers cannot conspire to mimic entanglement by classical means.  Buscemi concludes, as per the title of his paper, that all entangled quantum states are nonlocal \cite{buscemi}.  It is important to note in this regard that `nonlocal', as used by Buscemi, does not correspond to the usual sense of there being no local hidden variable model for a given quantum state \cite{bell}. Indeed, it is well known that some entangled states have precisely such a model \cite{werner}. Rather, `nonlocal' in Buscemi's sense is relative to complete trust by the referee in quantum randomness. 

Hence, while Buscemi's paper provides an important new method of witnessing entanglement, it is critical to examine the role played by trust in comparison to previous methods.  The aim of this paper is to show that the significance of Buscemi's semiquantum games, in this respect, is that while the referee needs to trust quantum mechanics and his or her own quantum state preparation devices, the referee does not need to trust the distant parties. Thus, we can think of his result as implying the existence of {\it quantum-refereed entanglement tests}. Further, we can generalize Buscemi's result by noting that, by modifying the type of randomness injected by the referee, the games can embody a lesser degree of trust in quantum mechanics by the referee, and thereby test a stronger degree of nonlocality. 

\begin{figure}[h] 
   \centering
   \includegraphics[width=6cm]{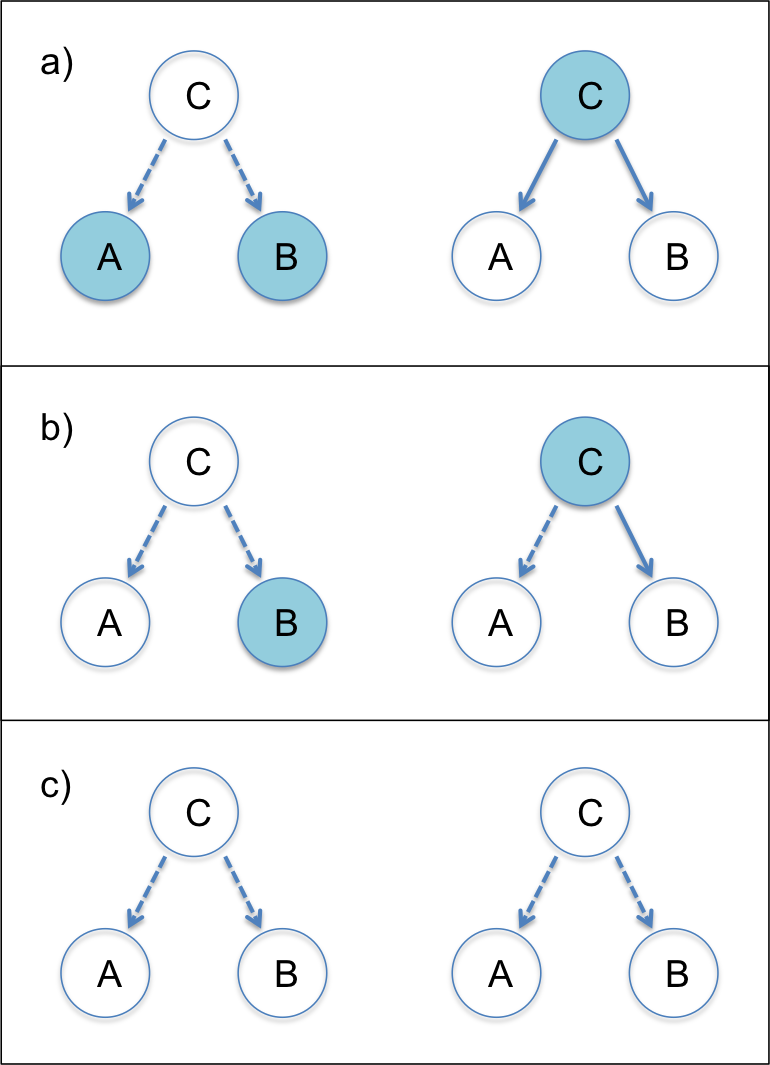} 
   \caption{Quantum-refereed games replace trust by quantum channels. Shaded regions indicate trust by the referee, Charlie, in the indicated party and in the quantum mechanical description of his or her apparatus. (Charlie's trust in himself is only indicated when he has a quantum state preparation device which he must trust.)  Dashed lines represent classical channels and full lines represent quantum channels. The first column represents conventional nonlocality tests with classical channels, with trust in the appropriate parties as introduced in Refs.~\cite{Wiseman2007,jones}; the second column represents the corresponding quantum-refereed tests. Trust in Alice and/or Bob can thus be substituted by trust in Charlie's state preparation devices and quantum channels to the corresponding party, to demonstrate a) entanglement and b) EPR-steering. Demonstration of c) Bell-nonlocality is unaltered as no party is assumed to be trusted.}
   \label{fig:trust_flow}
\end{figure}

In particular, we show that the nature of the states needed to be sent by the referee depends on the type of entangled resource that the observers claim to share, as indicated in Figure~\ref{fig:trust_flow}. The strongest type of nonlocality available through quantum mechanics, Bell inequality violations \cite{bell}, only requires the referee to provide classical randomness, corresponding to two classical channels. The strictly weaker nonlocal phenomenon of steering \cite{Wiseman2007,jones} (violating an EPR-steering inequality \cite{Cavalcanti2009})  requires one quantum channel and one classical channel. The still weaker phenomenon of entanglement witnessing (violating a separability inequality) \cite{ent_witnesses}, requires two quantum channels, as considered by Buscemi. The protocols in Figure~1 are very similar to those considered in Jones {\it et al.}~\cite{jones}, where only classical communication channels were used, together with varying degrees of trust by the referee of the observers; here,  such  `trust' is replaced by one-way quantum communication channels.
 
This paper is arranged as follows. In Section~\ref{sec:semiquantum} we review the results of \cite{buscemi}, showing how for every entangled state there exists a `semiquantum nonlocal game' that witnesses the entanglement. In Section~\ref{sec:are_all?} we discuss the foundational implications of that result, and in particular in what sense it can be said that ``all entangled states are nonlocal''. In Section~\ref{sec:hierarchy} we review the nonlocality hierarchy of entanglement, EPR-steering and Bell-nonlocality, and how this has previously been formulated in terms of trust by a referee in various parties during performance of an entanglement-verification task \cite{jones, Cavalcanti2009, Cavalcanti2011}. In Section~\ref{sec:replacing} we show how `semiquantum games' mandate a revision in the required degrees  of trust: with quantum-refereed games, trust from a referee in a distant party can be replaced by the combination of i) trust by the referee on his own preparation device and ii) a quantum channel to the untrustworthy parties. We show how this is also the case in the asymmetric task of EPR-steering~\cite{Wiseman2007,jones, Cavalcanti2009}. In Section~\ref{sec:conclusions} we offer some concluding remarks. 


\section{Semiquantum games and entanglement witnesses} \label{sec:semiquantum}

The work of Buscemi \cite{buscemi} centers on demonstrating the equivalence of two orderings on bipartite quantum states.  While these orderings are not of primary interest {\it per se} for our observations, they and their equivalence are briefly reviewed here to introduce Buscemi's `semiquantum nonlocal game' protocol, and to clarify the close connection of the latter with entanglement witnesses.

\subsection{LOSR ordering of states}

The first ordering on bipartite states is defined  via the class of completely positive maps corresponding to local operations and shared randomness (LOSR) \cite{buscemi,wolf}.  Such maps have the form $\sum_j \nu_j \,{\cal E}_j\otimes {\cal F}_j$, where $\{ \nu_j\}$ is a classical probability distribution and ${\cal E}_j$ and ${\cal F}_j$ are arbitrary completely-positive trace-preserving maps on the respective components of the state.  Two separated observers can generate an LOSR map on a shared bipartite state, without using any communication, via shared classical randomness corresponding to the distribution $\{\nu_j\}$.  It is clear that the class of LOSR maps forms a subset of the class of maps defined by local operations and classical communication (LOCC), since one can share any probability distribution ${\nu_j}$ by sampling it and communicating the result(s) classically. 

A state $\rho$ is defined to be LOSR sufficient for another bipartite state $\sigma$, written $\rho\rightarrowtail\sigma$, if and only if $\rho$ can be converted to $\sigma$ via some LOSR map. Thus,
\begin{equation} \label{losr}  
	\rho \rightarrowtail\sigma ~~{\rm if~ and~only~if}~~\sigma = \sum_j \nu_j ({\cal E}_j\otimes {\cal F}_j)(\rho)
\end{equation}
for some set $\{ \nu_j, {\cal E}_j, {\cal F}_j \}$.  It is straightforward to show that $\rho\rightarrowtail\sigma_{\rm sep}$ for any separable state $\sigma_{\rm sep}$, and that the converse, $\sigma_{\rm sep}\rightarrowtail \rho$ , only holds if $\rho$ is also separable \cite{buscemi}.  Thus the separable states form an equivalence class of bipartite states with respect to this ordering \cite{footpure}.

\subsection{`Semiquantum game' ordering of states}

The second ordering is defined via the class of `semiquantum nonlocal games', which we will abbreviate here to `semiquantum games'. These are games involving two experimenters, Alice and Bob, sharing a bipartite state $\rho_{AB}$, and a referee, Charlie, who will question them and receive answers, which he uses to calculate their payoff. The term `semiquantum', introduced by Buscemi \cite{buscemi}, refers to the fact that while Alice's and Bob's answers are of the usual (classical) sort, Charlie's questions are actually quantum states.

To formalise the game protocol, Charlie (i) sends state $\tau^s$ to Alice (representing a ``question'' labelled $s$), encoded on a subsystem $A_0$, with probability $p(s)$, and (ii) sends state $\omega^t$ to Bob (representing a ``question'' labelled $t$), encoded on a subsystem $B_0$,  with probability $q(t)$.  It is assumed that Alice and Bob cannot communicate with each other, although they may have agreed on a strategy beforehand. For each question asked by Charlie, Alice and Bob must transmit their respective answers, $x$ and $y$, back to Charlie. These answers may have been produced by Alice and Bob each carrying out a local joint measurement on the subsystem they received from Charlie and their component of $\rho_{AB}$, but no such assumption is made by Charlie. Their goal is to maximise the average value of some pre-agreed payoff function, $ \wp(x,y,s,t)$.  This protocol defines a semiquantum game, $G_{\rm sq}$ \cite{buscemi}. The quantum states Charlie uses to encode the questions are, in general, not mutually orthogonal, while the answers from Alice and Bob, being classical, could be represented by mutually orthogonal quantum states. 

Without loss of generality we can represent the measurements of Alice and Bob as positive operator valued measures (POVMs) $P=\{P_{A_0A}^x\}$ and $Q=\{Q_{BB_0}^y\}$, acting on $AA_0$ and $BB_0$, respectively. The maximum average payoff for a given shared state $\rho_{AB}$ and game $G_{\rm sq}$ is then given by
\[ {\wp}^{\star}(\rho_{AB},G_{\rm sq}) :=\max_{P,Q} \sum_{s,t,x,y}p(s)q(t)\mu(x,y|s,t)\,{\wp}(x,y,s,t), \]
where 
\begin{equation}
	\mu(x,y|s,t):={\rm Tr}\left[(P^x_{A_0A}\otimes Q^y_{BB_0})\,(\tau^s_{A_0}\otimes \rho_{AB}\otimes 
	\omega^t_{B_0})\right].
\nonumber
\end{equation}
It is natural to define a bipartite state $\rho$ to have greater utility for semiquantum games than bipartite state $\sigma$, written $\rho\succeq_{\rm sq}\sigma$, if the maximum average payoff for $\rho$ is always at least as much as that for $\sigma$, for any semiquantum game, i.e., 
\begin{equation} \label{sq}
	\rho\succeq_{\rm sq}\sigma~~{\rm if~ and~only~if}~~ \wp^{\star}(\rho,G_{\rm sq}) \geq  \wp^{\star}(\sigma,G_{\rm sq}) ~\forall\, G_{\rm sq}.
\end{equation}

\subsection{Entanglement witnesses via equivalence \\of orderings}

The main result of Buscemi is that the two orderings, $\rho \rightarrowtail\sigma$ and $\rho\succeq_{\rm sq}\sigma$, in Eqs.~(\ref{losr}) and (\ref{sq}), are equivalent \cite{buscemi}.  An immediate corollary is that any equivalence class of bipartite states with respect to LOSR maps is also an equivalence class with respect to semiquantum games.  That is, defining $\rho\leftarrowtail\!\rightarrowtail \sigma$ if and only if $\rho$ and $\sigma$ can be obtained from each other via suitable LOSR maps, one has
\[  
\rho\leftarrowtail\!\rightarrowtail \sigma~~{\rm if~ and~only~if}~~ \wp^{\star}(\rho,G_{\rm sq}) =  \wp^{\star}(\sigma,G_{\rm sq}) ~\forall\, G_{\rm sq} .
\]
As noted earlier, the set of separable states forms an equivalence class. Hence, it follows, in particular, that all separable states must yield the same maximum average payoff, $ \wp^{\star}_{\rm sep}(G_{\rm sq})$, for any game $G_{\rm sq}$, i.e., 
\begin{equation} \label{maxsep}
	\wp^{\star}(\sigma_{\rm sep},G_{\rm sq}) =  \wp^{\star}_{\rm sep}(G_{\rm sq})
\end{equation} 
for any separable state $\sigma_{\rm sep}$.  It is this result that leads to the direct connection of semiquantum games with entanglement witnesses.

In particular, it follows from Eq.~(\ref{maxsep}) that a given bipartite state $\rho$ is entangled if there is some corresponding semiquantum game $G_{\rm sq}$ such that
$
 \wp^{\star}(\rho,G_{\rm sq}) >  \wp^{\star}_{\rm sep}(G_{\rm sq}).
$
Hence, if $ \wp_{PQ }(\rho,G_{\rm sq})$ denotes the average payoff for this game when Alice and Bob make measurements corresponding to POVMs $P$ and $Q$, respectively, then this average payoff is an entanglement witness for $\rho$ whenever 
\begin{equation} \label{wit}
	\wp_{PQ}(\rho,G_{\rm sq})>  \wp^{\star}_{\rm sep}(G_{\rm sq}) .
\end{equation} 
Because the use of quantum states by Charlie obviates his need to trust Alice and Bob, we call such a protocol a {\it quantum-refereed entanglement test}. 
It is important to note that Alice and Bob do not need to make optimal choices of $P$ and $Q$ to verify that $\rho$ is entangled, as long as they achieve an average payoff for this game that is strictly larger than the maximum possible payoff for any separable state.

The equivalence between the LOSR ordering and the ordering induced by semiquantum games further implies that a given bipartite state $\rho$ is entangled {\it only if} there is some corresponding semiquantum game $G_{\rm sq}$ such that $\wp^{\star}(\rho,G_{\rm sq}) >  \wp^{\star}_{\rm sep}(G_{\rm sq})$. The proof is as follows. Suppose that a given entangled state $\rho$ has a maximum payoff $\wp^{\star}(\rho,G_{\rm sq})$ that is not larger than the maximum payoff $\wp^{\star}_{\rm sep}(G_{\rm sq})$ for separable states, for all $G_{\rm sq}$. From the equivalence between LOSR and semiquantum-game orderings, it thus follows that $\rho$ can be obtained by LOSR from a separable state. But LOSR transformations cannot create entanglement, and hence the assumption must be false. Thus, every entangled state must have a maximum payoff larger than that for separable states, for some semiquantum game.


\section{Are all entangled quantum states nonlocal?}
\label{sec:are_all?}

We now return to the title claim of Buscemi's paper, i.e., that ``all entangled quantum states are nonlocal''. The title and some of the introduction may suggest Buscemi is referring to Bell nonlocality \cite{bell}, but clearly this cannot be the case, as it would imply no local hidden-variable (LHV) model could explain the correlations of {\em any} entangled state. In particular,  it is well known that there are entangled states which have an explicit LHV model for all possible measurement schemes \cite{werner}.  Indeed, in the concluding paragraph of Buscemi's paper, he actually equates ``nonlocality'' with the property that the state has a greater utility than any separable state for semiquantum games \cite{buscemi}. By his own results, as described in Sec.~II above, this is in fact equivalent to equating ``nonlocality'' with ``entanglement''.

Now, all entangled states are obviously nonlocal in the sense of not being describable by a separable quantum state, and it is well known that for every entangled state there exists a witness, based on correlations between local measurements, that will detect the entanglement. The result of \cite{buscemi} would be trivial if this is all it were saying. The correct interpretation is a bit more subtle. It is that prior to this result, it was believed that to construct an appropriate entanglement witness protocol, it was necessary to \emph{trust} Alice and Bob to perform the appropriate measurements at their local systems. The result of Buscemi lifts the need for this trust on the distant parties, making it possible for entanglement to be detectable for all entangled states, even in an adversarial nonlocal game.

However, since Charlie encodes the questions in nonorthogonal states, he still needs to trust the devices used to prepare those states, and more importantly, he needs to trust quantum theory in the sense that Alice and Bob should not be able to distinguish those non-orthogonal states any more than is allowed by quantum theory. If those states were distinguishable in a theory superseding quantum mechanics, then Alice and Bob could obtain information about which question was encoded by Charlie in each run, and the situation reverts to that where the questions are encoded in orthogonal states. Or rather, the semiquantum nonlocal game protocol does not rule out the possibility that some local hidden-variables in the question-systems $A_0$ and $B_0$ may carry instructions for the devices of Alice and Bob that would be sufficient for them to `win' the game even when they share a separable state.

In other words, while the proof of \cite{buscemi} is device-independent for the distant parties, it is not fully device-independent as it requires trust in the preparation device of the referee, and it is not theory-independent, unlike the case for the proof of nonlocality allowed by a violation of a Bell inequality. The deep significance of Buscemi's result lies, therefore, not in a broader concept of ``nonlocality'' that subsumes the notion of entanglement, but rather in the removal of the need for a third party to trust Alice and Bob when verifying they share an entangled state --- even when there is an LHV model for this state.


\section{Trust and the Nonlocality Hierarchy} \label{sec:hierarchy}

While the result of \cite{buscemi} does not have the foundational implication that one might assume from its title, it does carry a  significant innovation in allowing the trust  by the referee in the distant parties to be replaced by i) trust in a local preparation device and ii) one-way quantum channels from the referee to the parties. This could have important implications for quantum communication. In Sec.~\ref{sec:replacing} we  further contextualize Buscemi's result in this setting by showing how his proof can be generalized to quantum-refereed EPR-steering games.  These require a one-way quantum channel on only one of two sides. In particular, we show that all steerable states give an advantage over all non-steerable states in such games. The implication is that the formalism of \cite{buscemi} mandates a revision of the existing definitions of the three nonlocality classes of entanglement, EPR steering and Bell nonlocality in terms of trust, which were introduced in \cite{Wiseman2007} and  further developed in Refs.~\cite{jones,Cavalcanti2009,Cavalcanti2011}. We now describe them briefly, with special attention to the notion of steering. 

In Ref.~\citep{jones} the nonlocality hierarchy of entanglement, steering and Bell nonlocality was presented in terms of a task wherein  a referee, Charlie, wants to verify that Alice and Bob share an entangled state. Alice and Bob share a number of copies of a bipartite state $\rho_{AB}$, and for each of those Charlie will ask them to perform one of a number of measurements chosen by Charlie at random. If Charlie trusts Alice and Bob (and their apparatuses), then it will be sufficient for Alice and Bob to violate a separability criterion; if Charlie trusts neither Alice nor Bob, then they will need to violate a Bell inequality; if Charlie trusts Bob (say), but not Alice, then they will need to demonstrate that Alice can steer Bob's system.

Let's consider the task of steering, or EPR-steering as this task is denoted in \cite{Cavalcanti2009} (where the concept of EPR-steering inequalities was defined). The term refers to a phenomenon identified by Schr\"{o}dinger in a 1935 article \cite{Sch35}, motivated by the effect discovered by Einstein, Podolsky and Rosen  in their seminal paper in that same year \cite{EPR35}. Schr\"{o}dinger noted that it was ``rather discomforting'' that, despite having no direct access to Bob's system, Alice was able, by different choices of measurement on her system, labelled by $s \in \mathcal{S}$, to ``steer'' Bob's state into states from distinct ensembles $E^s \equiv \{\tilde{\rho}_B^{x|s}:x\in{\cal X}_s\}$. Here ${\cal X}_s$ denotes the set of possible measurement outcomes, and the tilde indicates that those states are unnormalized, their trace being the probability of outcome $x$ given $s$.

In the EPR-steering task, Alice must be able to prove that she is able to steer Bob's system. This is impossible if, and only if, for all $s \in {\cal S}$ and for all $x\in{\cal X}_{s}$ there exists an ensemble $\{p(\xi)\rho_{B}^{\xi} :\xi \in \Xi\}$ of local quantum states for Bob's system and a stochastic map $p(x|s,\xi)$ such that
\begin{equation}\label{LHS}
	\tilde{\rho}_B^{x|s}=\sum_{\xi \in \Xi} p(x|s,\xi)p(\xi) \rho_{B}^{\xi} 
\end{equation}
In other words, Alice can succeed if and only if the reduced ensembles for Bob cannot be generated by a model wherein Bob has a local quantum state that is classically correlated with some variables in Alice's possession. To use the terminology of Refs.~\cite{Wiseman2007,jones}, steering is demonstrated if and only if their correlations cannot be explained in terms of a \emph{local hidden state model} for Bob.


\section{Replacing trust by quantum randomness} \label{sec:replacing}

As mooted in Sec.~\ref{sec:hierarchy}, we generalize Buscemi's semiquantum games by noting that the type of randomness input to each observer by the referee naturally falls into one of two types:  classical randomness, corresponding to sending classical signals to the observer (represented by mutually orthogonal input states in the protocol); and quantum randomness, corresponding to sending arbitrary quantum signals to the observer. These categories may be represented by classical and quantum communication channels, respectively.  Hence, the types of nonlocality verified via semiquantum games naturally fall into one of three categories, depending on whether two classical channels,  one quantum and one classical channel, or two quantum channels, are required by the referee.  It is shown below that these categories correspond to verifying one of Bell nonlocality, EPR-steering, or entanglement, respectively, as per Figure~\ref{fig:trust_flow}.

\subsection{Quantum-refereed verification of entanglement }

We begin by explaining in more detail how semiquantum games allow a referee, Charlie, to verify that Alice and Bob share some entangled state, without Charlie having to trust either of Alice and Bob.  First, Alice and Bob may choose to tell Charlie the density operator, $\rho_{AB}$, describing their state. This is not an essential part of the protocol, but that would be in their best interest for the purposes of verifying entanglement, as it would allow Charlie to optimise the choice of game for their particular state. The referee then chooses a suitable semiquantum game, $G_{\rm sq}$, corresponding to suitable input ensembles $\{\tau^s;p(s)\}$ and $\{\omega^t;q(t)\}$ and payoff function $ \wp(x,y,s,t)$, and informs Alice and Bob of his choice.  They are allowed to consult on a strategy to decide on suitable POVMs $P$ and $Q$, and from then on cannot communicate with each other.  The game is then played, and entanglement is verified by Charlie if the average payoff is larger than the maximum separable payoff, $ \wp^{\star}_{\rm sep}(G_{\rm sq})$, for that game, as per Eq.~(\ref{wit}).

Note that even if Alice and Bob use measurements and/or shared randomness --- for example, to discriminate as well as possible between nonorthogonal states sent by Charlie --- this does not allow them to cheat in the above protocol. No such strategy can make their shared state appear more entangled than it is. In particular, any cheating strategy corresponds to applying some LOSR map $\sum_j\nu_j\,{\cal E}_j\otimes {\cal F}_j$ to the joint state  $\tau^s_{A_0} \otimes \rho_{AB}\otimes \omega^t_{B_0} $, before the measurement of POVMs $P$ and $Q$.  However, this is formally equivalent to instead making a joint measurement, on $\tau^s_{A_0} \otimes\rho_{AB}\otimes \omega^t_{B_0} $, corresponding to the POVM $\{J^{xy}\}$, with $J^{xy}= \sum_j \nu_j\, {\cal E}^d_j(P^x)\otimes {\cal F}_j^d(Q^y)$, where $\phi^d$ denotes the dual of map $\phi$ \cite{dual}.  Hence, the strategy yields an average payoff given by the mean value, $\sum_j \mu_j\, \wp_j$, of the average payoffs $ \wp_j$ corresponding to Alice and Bob measuring the POVMs $\{{\cal E}^d_j(P^x)\}$ and $\{{\cal F}_j^d(Q^y)\}$. Since this mean value cannot be greater than the largest value of $ \wp_j$, which in turn is bounded by $ \wp^{\star}(\rho_{AB},G_{\rm sq})$, Alice and Bob cannot use any such strategy to obtain a higher average payoff than is possible for $\rho_{AB}$. 

Thus, Charlie can be sure, as long as Alice and Bob cannot communicate during the game, that an average payoff greater than $ \wp_{\rm sep}^{\star}(G)$, as per Eq.~(\ref{wit}), is a genuine signature of entanglement. The referee does not have to know what Alice and Bob actually do,  or or trust what they say --- knowing that they are limited to LOSR maps and local measurements, Alice and Bob can be treated purely as black boxes into which quantum states are input by Charlie, and measurement results are output to Charlie, permitting calculation of the average payoff.  In this way quantum-refereed games can witness entanglement with no trust of Alice and Bob by Charlie.

The mechanism that allows such lack of trust is the injection of sufficient randomness by Charlie, via the use of suitable input ensembles $\{\tau^s;p(s)\}$ and $\{\omega^t;q(t)\}$. This randomness negates the possibility of conspiracies between Alice and Bob of the type mentioned in Sec.~\ref{Sec:Intro}, and more generally the possibility of cheating via any LOSR map as above. We note here that Charlie could use entangled states instead of quantum channels, to allow him to make (random) choices of signals {\it after} Alice and Bob have sent their measurement results, or, alternatively, in a region that is space-like separated from Alice and Bob. In this case Charlie sends half of an entangled state to Alice (or Bob), and makes a measurement on his half. The measurement result establishes $s$ (or $t$) for Charlie, and is equivalent to sending the reduced state $\omega_s$ (or $\tau_t$) of the other component to Alice (or Bob), with probability $p(s)$ (or $q(t)$).

\subsection{Verification of Bell nonlocality} \label{sub:Bell}

We say a state is Bell nonlocal, if it violates some Bell inequality \cite{bell}. To test for Bell inequality violation, without trusting Alice and Bob, Charlie only needs to send classical signals $s$ and $t$ to Alice and Bob, respectively, via classical communication channels (or, equivalently, states from mutual orthogonal sets $\{\pi_{A_0}^s\}$ and $\{\pi_{B_0}^t\}$). The corresponding games may be called Bell games, or nonlocal games \cite{buscemi,wat,silman}. Since only classical signals are sent and received, Charlie need not even trust quantum mechanics to verify Bell nonlocality. 

As shown by Buscemi, for any state that enables the violation of some Bell inequality, there is always a suitable Bell game for which that state will perform better than all unentangled states\cite{buscemi,silman}.  To make this clear, assume that Alice and Bob claim their correlations violate the Bell inequality 
\begin{equation}
	\sum_{x,y,s,t} w(x,y,s,t)p(x,y|s,t)\leq B,
\end{equation}
where the upper bound $B$ holds for any correlations modeled by a local hidden variable theory \cite{bell} (and in particular for any separable state). A corresponding Bell game, $G_{\rm Bell}$, is defined by Charlie sending signals $s$ and $t$ with two arbitrary non-vanishing probability distributions $p(s)$ and $q(t)$, respectively, and choosing the payoff function $ \wp(x,y,s,t)=w(x,y,s,t)/[p(s)\,q(t)]$. 

Alice and Bob could produce correlations violating the inequality by sharing an appropriate Bell-nonlocal state $\rho_{AB}$, and through appropriate choices of POVMs $\{P_A^x(s)\}$ and $\{Q_B^y(t)\}$ such that $p(x,y|s,t)={\rm Tr}[(P_A^x(s)\otimes Q_A^y(t))\rho_{AB}]$ (or equivalently, through appropriate choices of POVMs $\{P_{A_0A}^x\}$ and $\{Q_{BB_0}^y\}$ such that $p(x,y|s,t)={\rm Tr}[(P_{A_0A}^x \otimes Q_{BB_0}^y)(\pi_{A_0}^s\otimes\rho_{AB}\otimes\pi_{B_0}^t)]$). However, no such assumption is made by Charlie about the mechanism by which Alice and Bob produce their results. Thus a violation of a Bell inequality demonstrates, in a theory-independent way, that the correlations produced by Alice and Bob cannot be explained locally.

\subsection{Quantum-refereed verification of EPR-steering}

To derive the existence of quantum-refereed verification of EPR-steering, a nontrivial modification of Buscemi's proof for entanglement verification is required.  It is nontrivial because the relevant class of maps is no longer the LOSR class (see Sec.~II), but the class of `local operations with steering  and shared randomness'; a suitable `steerability' ordering must be defined on bipartite states; and the role of separable states must be replaced by non-steerable states.  However, when these concepts are suitably defined, it becomes relatively straightforward to apply the methods of Ref.~\cite{buscemi} to allow verification of EPR-steering in the case where the referee does not trust Alice or Bob.

\subsubsection{LOSSR ordering of bipartite states}

As per Sec.~IV, Alice steers Bob's component of a shared bipartite state $\rho$, to the ensemble $E^s=\{\tilde{\rho}^{x|s}_B\}$, by carrying out a local measurement labelled by $s$ with outcomes labelled by $x$.  If the POVM element corresponding to Alice's outcome $x$, given measurement $s$, is denoted by $P^{x|s}_A$, and the corresponding measurement operation by ${\cal E}^{x|s}$, then
\begin{equation} \label{inter}
	\tilde{\rho}^{x|s}_B = {\rm Tr}_A [(P^{x|s}_A\otimes \id_B)\,\rho] = {\rm Tr}_A  [ ({\cal E}^{x|s}\otimes I_B)(\rho)] ,
\end{equation}
where $\id_B$ and $I_B$ denote, respectively, the identity operator and identity operation for Bob's component.  Noting that Bob can also operate locally on $E^s$, by  applying any completely-positive trace-preserving map ${\cal F}$, and that Alice could choose, for a given value of the label $s$, to measure some POVM $\{ P_A^{x|s}(i)\}$ with probability $\nu(i)$, it follows that with shared randomness they can steer Bob's component to any ensemble $\{\tilde{\rho}^{x|s}_B\}$ such that 
\begin{equation} \label{LOSSR} 
	\tilde{\rho}^{x|s}_B =\phi^{x|s}(\rho)  :=\sum_i \nu(i){\rm Tr}_A [({\cal E}^{x|s}_i\otimes {\cal F}_i)(\rho)].
\end{equation}
We will refer to such maps $\phi^{x|s}$ as `local operations with steering and shared randomness' (LOSSR) maps. The set of ensembles that Bob's component can be steered to via LOSSR maps will be denoted by $S_B(\rho)$.

It is natural to define bipartite state $\rho$ to be LOSSR-sufficient for bipartite state $\sigma$, written $\rho\rightarrowtail_{ \rm st}\sigma$, if and only if Bob's component of $\rho$ can be steered to any ensemble that Bob's component of $\sigma$ can be steered to, via LOSSR maps, i.e.,
\begin{equation} 
	\rho\rightarrowtail_{ \rm st}\sigma {\rm ~~if~and~only~if~~} S_B(\sigma)\subseteq S_B(\rho) .
\end{equation}
As might be expected, $\rho \rightarrowtail_{ \rm st}\sigma_{\rm nst}$ for any non-steerable state $\sigma_{\rm nst}$.  This is because the correlations of any non-steerable state  $\sigma_{\rm nst}$, i.e., any state admitting a local hidden state model of the form of Eq.~(\ref{LHS}), can be given a classical-quantum model~\cite{discord} $\sigma_{\rm cq}=\sum_\xi p(\xi) \pi^\xi_\alpha \otimes \rho^\xi_B$, where the $\pi^\xi_\alpha$ are orthogonal  projectors on a sufficiently enlarged Hilbert space $H_\alpha$.   Since $\sigma_{\rm cq}$  is a separable state, it can be generated {\it ab initio} via shared randomness from any state $\rho$ via  a `discard and prepare' LOSR map (with $\nu( \xi)\equiv p(\xi)$ and $ ({\cal E_\xi}\otimes{\cal F_\xi})(\rho):=\pi_\xi\otimes \rho^\xi_B$). 
Defining the POVM $\{P_\alpha^{x|s}\}$ via  $P_\alpha^{x|s}:=\sum_\xi p(x|s,\xi)\pi^\xi_\alpha$, then  $\tilde{\sigma}_{{\rm cq},B}^{x|s}={\rm Tr}_\alpha[(P_\alpha^{x|s}\otimes\id_B)\sigma_{\rm cq}]=\sum_\xi p(\xi)p(x|s,\xi)\rho_B^\xi$, thus reproducing, via LOSSR,  the steering ensembles that Alice can prepare for Bob from $\sigma_{\rm nst}$. It follows immediately that  $S_B(\sigma_{\rm nst})=S_B(\sigma_{\rm cq}) \subseteq S_B(\rho)  $. Hence, $\rho \rightarrowtail_{ \rm st}\sigma_{\rm nst}$ as claimed.   

Conversely, no steerable state $\rho$ can be obtained from a non-steerable state $\sigma_{\rm nst}$ via an LOSSR map (since such maps, as per Eqs.~(\ref{inter}) and (\ref{LOSSR}), preserve the existence of a local hidden state model as per Eq.~(\ref{LHS}) for Bob's ensembles).  It immediately follows that the non-steerable states form an equivalence class with respect to $\rightarrowtail_{\rm st }$, which lies at the bottom end of the ordering. Note that since the set of separable states is a proper subset of the set of non-steerable states \cite{Wiseman2007,jones}, it follows that the LOSSR ordering is weaker than the LOSR ordering defined by Buscemi (see Sec.~II above).

\subsubsection{Steering-game ordering of bipartite states}

We now define quantum-refereed steering games. These games are a hybrid of nonlocal games (Bell nonlocality  games) and Buscemi's semiquantum games (quantum-refereed entanglement verification). In the case of steering verification, Charlie picks with probability $p(s)$ a question $s\in\mathcal{S}$ for Alice and encodes it in a state from an orthonormal set $\pi=\{\pi^s\}$, and with probability $q(t)$ a question $t\in\mathcal{T}$ for Bob and encodes it in a state from an arbitrary set of normalized states $\omega=\{\omega^t\}$. As before, Alice and Bob must compute their answers $x\in\mathcal{X}$ and $y\in\mathcal{Y}$ without communicating with each other, but may share an arbitrary bipartite quantum state $\rho_{AB}$. For each combination of $(s,t,x,y)$ they receive a payoff $\wp(s,t,x,y)$.  
 
The maximum average payoff which can be obtained for state $\rho$, via a given quantum-refereed steering game $G_{\rm st}$, will be denoted by $\wp^{\star}(\rho,G_{\rm st})$.  It is natural to define state $\rho$ to have greater utility for such games than state $\sigma$, written $\rho\succeq_{\rm st}\sigma$, if and only if the maximum payoff for $\rho$ is always at least as great as the maximum payoff for $\sigma$, i.e.,
\begin{equation}
	\rho\succeq_{\rm st}\sigma {\rm ~~if~and~only~if~~}\wp^{\star}(\rho,G_{\rm st}) \geq \wp^{\star}(\sigma,G_{\rm st})~\forall\, G_{\rm st}.
\end{equation}
Note the close analogy with the ordering $\succeq_{\rm sq}$ in Eq.~(\ref{sq}).

\subsubsection{Equivalence of orderings and steering witnesses}

With the above definitions, it is now possible to prove the equivalence of the two orderings, i.e., 
\begin{equation} \label{stequiv}
\rho\rightarrowtail_{\rm st}\sigma {\rm ~~if~and~only~if~~}\rho\succeq_{\rm st}\sigma ,
\end{equation}
via a suitable modification of the derivation in the supplemental material of Ref.~\cite{buscemi}.  Indeed, the asymmetry between Alice and Bob in quantum-refereed steering games leads to some simplifications relative to the semiquantum game case.  The details are given in the Appendix.
 
Since the set of non-steerable states form an equivalence class with respect to $\rho\rightarrowtail_{\rm st}$, it immediately follows that they form an equivalence class with respect to $\succeq_{\rm st}$, and hence that the maximum payoff function of any quantum-refereed steering game $G_{\rm st}$ has the same value $\wp_{\rm nst}^{\star}(G_{\rm st})$ for any non-steerable state $\sigma_{\rm nst}$, i.e.,
\begin{equation}
\wp^{\star}(\sigma_{\rm nst}, G_{\rm st}) = \wp_{\rm nst}^{\star}(G_{\rm st}).
\end{equation}
Further, for any steerable state $\rho$, there must be at least one quantum-refereed steering game $G_{\rm st}$ such that
\begin{equation}
\wp^{\star}(\rho, G_{\rm st}) > \wp_{\rm nst}^{\star}(G_{\rm st}).
\end{equation}
Such a game provides a witness to Alice and Bob sharing a steerable bipartite state, in direct analogy to the role of semiquantum games witnessing sharing of an entangled state (see Sec.~II).  Note that unlike the semiquantum game scenario, Charlie only needs one quantum channel to witness steering, when neither Alice nor Bob are trusted, as illustrated in Fig.~\ref{fig:trust_flow}.


\section{Conclusions} \label{sec:conclusions}

In this paper we have analyzed and generalized recent work by Buscemi~\cite{buscemi}. Our analysis shows that the foundational implication of that work needs to be carefully considered --- it does not imply that all entangled states are Bell-nonlocal as one may na\"{i}vely suspect from the title-claim. On the other hand, Buscemi's work  carries interesting implications for quantum communications, as it allows, through the use of quantum channels from the referee to the distant parties in a nonlocality test, for the substitution of trust in the distant parties by trust in the referee's quantum devices.  

Thus we can think of Charlie (the referee) suitably ``quantizing'' the instructions sent by him to Alice and Bob regarding their measurement settings, in order that he not have to trust them.  Adopting this viewpoint, we were able to generalize Buscemi's work from entanglement verification to EPR-steering verification. In the latter case, the transfer of trust takes place for only one party, via a quantum channel from the Charlie to (say) Bob, while the channel to Alice becomes classical. This defines a \textit{quantum-refereed steering game}. Analogously to the case of entanglement verification considered by Buscemi, we have shown that a state is steerable (i.e., it cannot be described by a local hidden state model~\cite{Wiseman2007}) if and only if there exists a quantum-refereed steering game for which it does better than all non-steerable states. 

It is important to note that our proof does not provide an explicit method for constructing a quantum-refereed game that will demonstrate EPR steering for a given steerable state, just as the proof of Ref.~\cite{buscemi} does not show how to construct a semiquantum nonlocal game for witnessing entanglement of a given entangled state. However, the construction of suitable quantum-refereed steering games should prove a tractable problem for highly symmetric states such as Werner states \cite{werner}, for which the hierarchy of entanglement is well understood \cite{jones}. Moreover, the details of the proof given in the appendix imply -- analogously to the case of Ref.~\cite{buscemi} -- that there is always a suitable game for verifying steerability in which Bob makes a generalized Bell-state measurement --- where it is experimentally feasible to implement such measurements for two-qubit states \cite{Gao2010}. Hence, it is expected that quantum-refereed steering games can be designed for specific states. 

For experimental implementations, it is noted that quantum-refereed EPR steering as we have introduced here would presumably be considerably easier to test than quantum-refereed entanglement witnessing as introduced by Buscemi. The reason is that the former only requires one joint measurement to be reliably made in each run, by Bob (over the portion of his shared system and the quantum input sent by Charlie), whereas the latter requires two joint measurements (i.e., by each one of Alice and Bob). Thus we expect the quantum-refereed EPR-steering we have introduced here to allow the first experimental application of Buscemi's idea of quantized measurement settings.
 
\acknowledgments{HMW and MJWH are supported by the ARC Centre of Excellence CE110001027. EGC is funded by an ARC DECRA DE120100559.}

\appendix

\subsection*{Note added}

After submission of this paper, a paper has appeared which gives a nice method for explicitly constructing loss-tolerant quantum-refereed nonlocal games \cite{branc}. The possibility of modifying this method, to allow an analogous construction of quantum-refereed steering games, is under investigation.

\section{Proof of Eq.~(\ref{stequiv})}
 
To prove the equivalence of orderings in Eq.~(\ref{stequiv}), we use methods and notation inspired by Buscemi's proof of the equivalence of the orderings in Eqs.~(1) and (2) \cite{buscemi}.

Consider first the case $\rho\longrightarrow_{\rm st}\sigma$.  Hence, $S_B(\sigma)\subseteq S_B(\rho)$, so that any steering ensembles generated from $\sigma$ can also be generated from $\rho$ via LOSSR. It is therefore trivial that the maximum payoff for $\rho$ must be at least as great as the maximum payoff for $\sigma$ for any quantum-refereed steering game, i.e., that  $\rho\succeq_{\rm st}\sigma$.

To prove the converse direction, define  $\tilde{\pi}^s:=p(s)\pi^s$ and $\tilde{\omega}^t:=q(t)\omega^t$. Consider a particular quantum-refereed steering verification game $G_{\rm st}$. The average payoff when Alice and Bob measure POVMs $P=\{P^x_{A_0A}:x\in\mathcal{X}\}$ and $Q=\{Q^y_{BB_0}:y\in\mathcal{Y}\}$, is given by
\begin{multline}\label{g_Gst}
	g(\rho_{AB};G_{\rm st};P,Q)=\\
	\sum_{s,t,x,y}\wp(s,t,x,y)\mu_{P,Q}(s,t,x,y),
\end{multline}
where
\begin{multline}\label{mustxy}
	\mu_{P,Q}(s,t,x,y)=\\
	\mathrm{Tr}[(P^x_{A_0A}\otimes Q^y_{BB_0})(\tilde{\pi}^s_{A_0}\otimes\rho_{AB}\otimes \tilde{\omega}^t_{B_0})].
\end{multline}

Imagine now the set of all probabilities $\mu_{P,Q}(s,t,x,y)$ that Alice and Bob can obtain by varying $P$ and $Q$ in \erf{mustxy}, and keeping $\rho_{AB}$, $\{\pi^s\}$ and $\{\omega^t\}$ fixed. This is not a convex set, since it is not the case   that, for all $w \in [0,1]$ and all local measurements $P', P'', Q', Q''$,  one can find other local measurements $P$ and $Q$ such that $w\mu_{P',Q'}(s,t,x,y)+(1-w)\mu_{P'',Q''}(s,t,x,y)  =\mu_{P,Q}(s,t,x,y)$. The reason is that in general $w(P'\otimes Q')+(1-w)(P''\otimes Q'')$ cannot be written as a product measurement $P\otimes Q$. As our proof relies on convexity properties, we extend this set by considering arbitrary convex combinations of local POVMs, $Z_{A_0ABB_0}^{x,y}:=\sum_i \nu(i)P^x_{A_0A}(i)\otimes Q^y_{BB_0}(i)$, where $\nu(i)$ are probabilities. Note however, that since \erf{g_Gst} is linear in $P\otimes Q$ (and thus, in particular, convex), and since a convex function on a convex set attains its maximum at the extreme points, the maximum payoff for a given {quantum-refereed steering verification} game $G_{\rm st}$ given a quantum state $\rho_{AB}$ is given by
\begin{equation}
	\wp^{*}(\rho_{AB};G_{\rm st}):=\mathrm{max}_{Z}\sum_{s,t,x,y}\wp(s,t,x,y)\mu_{Z}(x,y,s,t).
\end{equation}
We now assume that $\rho_{AB} \succeq_{\rm st} \sigma_{A'B'}$, i.e., that $\wp^{*}(\rho_{AB};G_{\rm st})\geq\wp^{*}(\sigma_{A'B'};G_{\rm st})$ for all {quantum-refereed steering verification} games $G_{\rm st}$. Following \cite{buscemi}, this implies that for any choice of $\mathcal{S}, \mathcal{T}, \mathcal{X}, \mathcal{Y}, A_0, B_0, \pi, \omega$, and for any POVMs $Z$, there exists a POVM $\bar{Z}$ such that
\begin{multline} \label{ZbarZ1}
	\mathrm{Tr}[\bar{Z}^{x,y}_{A_0ABB_0}(\tilde{\pi}^s_{A_0}\otimes\rho_{AB}\otimes \tilde{\omega}^t_{B_0})] =\\
	\mathrm{Tr}[Z^{x,y}_{A_0A'B'B_0}(\tilde{\pi}^s_{A_0}\otimes\sigma_{A'B'}\otimes \tilde{\omega}^t_{B_0})],
\end{multline}
for all $s,t,x,y$.
We now, on the referee's behalf, choose $A_0$ and $B_0$ to be such that $\mathbb{H}_{A_0}\equiv\mathbb{H}_{A'}$ and $\mathbb{H}_{B_0}\equiv\mathbb{H}_{B'}$, introduce a further auxiliary system $B_1$ with $\mathbb{H}_{B_1}\equiv\mathbb{H}_{B_0}$, and choose a particular set $\{\omega^t\}$ given by
\begin{equation}
	\omega^t_{B_0}=\mathrm{Tr}_{B_1}[(\id_{B_0} \otimes \Upsilon^t_{B_1}) \Psi^+_{B_0B_1}],
\end{equation}
where $\Psi^+$ denotes a maximally entangled state and $\Upsilon=\{\Upsilon^t: t \in \mathcal{T}\}$ is an informationally complete POVM. With this choice, we can rewrite \erf{ZbarZ1} as 
\begin{multline} \label{ZbarZ2init}
	\mathrm{Tr}[(\bar{Z}^{x,y}_{A_0ABB_0} \otimes \Upsilon^t_{B_1})(\tilde{\pi}^s_{A_0}\otimes\rho_{AB} \otimes \Psi^+_{B_0B_1})] =\\
	\mathrm{Tr}[(Z^{x,y}_{A_0A'B'B_0} \otimes \Upsilon^t_{B_1})(\tilde{\pi}^s_{A_0}\otimes\sigma_{A'B'}\otimes \Psi^+_{B_0B_1})],
\end{multline}
for all $s,t,x,y$.

Because $\Upsilon$ is informationally complete, we can simplify that expression by identifying the reduced states for $B_1$ on each side. Choosing now a product POVM on the right side, we obtain
\begin{multline}\label{ZbarZ2}
	\mathrm{Tr}_{A_0ABB_0}[(\bar{Z}^{x,y}_{A_0ABB_0} \otimes \id_{B_1})(\tilde{\pi}^s_{A_0}\otimes\rho_{AB} \otimes \Psi^+_{B_0B_1})] =\\
	\mathrm{Tr}_{A_0A'B'B_0}[(P^{x}_{A_0A'}\otimes Q^y_{B'B_0} \otimes \id_{B_1})(\tilde{\pi}^s_{A_0}\otimes\sigma_{A'B'}\otimes \Psi^+_{B_0B_1})],
\end{multline}
for all $s,x,y$. Note that this result is analogous to Eq.~(5) of the supplementary material for Ref.~\cite{buscemi}, but (due to the orthogonality of the states $\tilde{\pi}^s_{A_0}$), does not require the introduction an auxiliary system $A_1$, nor an informationally complete set $\Theta^s_{A_1}$, nor a Bell state $\Psi^+_{A_1A_0}$. 

We now choose the POVM $Q$ to be the generalized Bell measurement on $B'B_ 0$, and denote the right-hand side of \erf{ZbarZ2} by $\tilde{\sigma}^{s,x,y}_{B_1}$. Using the protocol of quantum teleportation, we can find unitary operators $V^y: \mathbb{H}_{B_1}\rightarrow\mathbb{H}_{B'}$ such that
\begin{equation}
	\sum_y (V^y_{B_1}) \tilde{\sigma}^{s,x,y}_{B_1} (V^y_{B_1})^{\dagger} = \tilde{\sigma}^{x|s}_{B'},
\end{equation}
where $\tilde{\sigma}^{x|s}_{B'} := \mathrm{Tr}_{A_0A'}[P^{x}_{A_0A'}(\tilde{\pi}^s_{A_0} \otimes \sigma_{A'B'})]$ is the (un-normalized) reduced state at Bob following an outcome $x$ for question $s$ at Alice's.

On the other hand, since $\rho_{AB} \succeq_{\rm st} \sigma_{A'B'}$, we know that there exists a POVM $\bar{Z}$ such that, according to \erf{ZbarZ2}
\begin{align}
	\tilde{\sigma}^{x|s}_{B'}   =   \sum_y V^y_{B_1} \mathrm{Tr}_{A_0ABB_0}[&(\bar{Z}^{x,y}_{A_0ABB_0} \otimes \id_{B_1}) \times \nn \\
	&  (\tilde{\pi}^s_{A_0}\otimes\rho_{AB} \otimes \Psi^+_{B_0B_1})] (V^y_{B_1})^\dagger. \nn
\end{align}
Expanding $\bar{Z}^{x,y}_{A_0ABB_0}=\sum_i \nu(i)\bar{P}^x_{A_0A}(i)\otimes\bar{Q}^y_{BB_0}(i)$, defining the POVM $\{\bar{P}^{x|s}_{A}(i)\}$ via
\[ \bar{P}^{x|s}_A(i):={\rm Tr}_{A_0}[\bar{P}^{x|s}_{A_0A}(i)\,(\pi^s_{A_0}\otimes \id_A)] , \]
 and defining the completely-positive trace-preserving map
\begin{multline}
	\mathcal{F}^i(w_B):= \\
	\sum_y V^y_{B_1} \mathrm{Tr}_{BB_0}[(\bar{Q}^{y}_{BB_0} \otimes \id_{B_1})(w_{B} \otimes \Psi^+_{B_0B_1})] (V^y_{B_1})^\dagger, \nn
\end{multline}
we conclude that 
\begin{multline}\label{conclude}
	\tilde{\sigma}^{x|s}_{B'}= 
	\sum_i \nu(i) \mathrm{Tr}_{A }[(\bar{P}^{x|s}_A(i) \otimes \id_B)\, (I_A \otimes \mathcal{F}^i_B)(\rho_{AB})]. 
\end{multline}
Comparing with Eqs.~(\ref{inter}) and (\ref{LOSSR}), it follows that any ensemble obtained from $\sigma_{A'B'}$ can also be obtained via an LOSSR map from $\rho_{AB}$, and hence that $\rho_{AB}\longrightarrow_{\rm st}\sigma_{A'B'}$ as required.

\bibliographystyle{unsrt}

\begin{thebibliography}{99}
\bibitem{footpure} It is also straightforward to show that any two {\it pure} bipartite states are equivalent, with respect to the LOSR ordering, if and only if they are related by local unitary transformations, i.e., if and only if they have the same Schmidt coefficients. This follows from Theorem 1 of Ref.~\cite{Nie99}, recalling that the class of LOSR maps is a subset of the class of LOCC maps.
\bibitem{dual} The dual $\phi^d$ of a linear map $\phi$ is defined by ${\rm Tr}[X\phi(Y)]={\rm Tr}[\phi^d(X)Y]$ for all trace-class operators $X$ and $Y$.
\bibitem{discord} For connection with other literature, note that this ``classical-quantum'' state is equivalent to a state with zero Alice-discord \cite{Zurek,Brukner}.
\end{thebibliography}

\end{document}